\documentclass[12pt,a4paper]{article}
\pdfoutput=1
\usepackage[T1]{fontenc}
\usepackage[sc,osf]{mathpazo}
\usepackage{a4wide}  
\usepackage{latexsym,amsthm,amsfonts,amsmath,mathrsfs,amssymb}
\usepackage[unicode,implicit]{hyperref}
\hypersetup{%
  pdftitle    = {Higher order gravities and the Strong Equivalence Principle}
  pdfkeywords = {curvature, higher order, energy-momentum tensor,
    self-coupling, consistency, gauge invariance, Noether identity},
  pdfauthor   = {Tomas Ortin},
  plainpages  = true,
  colorlinks  = true,
  citecolor   = blue,
  urlcolor    = red,
  linkcolor   = black
}
\newcommand{\hepth}[1]{{\tt
\href{http://www.arXiv.org/abs/hep-th/#1}{hep-th/#1}}}
\newcommand{\grqc}[1]{{\tt
\href{http://www.arXiv.org/abs/gr-qc/#1}{gr-qc/#1}}}

\newcommand{\arxiv}[1]{{\tt arXiv:\href{http://www.arXiv.org/abs/#1}{#1}}}
\newcommand{\doi}[1]{{\tt DOI:\href{http://dx.doi.org/#1}{#1}}}

\makeatletter
\@addtoreset{equation}{section}
\makeatother

\begin{document}

\begin{flushright}
\small
IFT-UAM/CSIC-17-032\\
\texttt{arXiv:1705.03495 [gr-qc]}\\
May 8\textsuperscript{th}, 2017\\
\normalsize
\end{flushright}

\vspace{1.5cm}

\begin{center}

{\LARGE {\bf {Higher order gravities\\[.5cm] and \\[.5cm] the Strong Equivalence Principle}}}
 
\vspace{1.5cm}

\renewcommand{\thefootnote}{\alph{footnote}}
{\sl\large Tom\'{a}s Ort\'{\i}n}\footnote{E-mail: {\tt Tomas.Ortin [at] csic.es}}

\setcounter{footnote}{0}
\renewcommand{\thefootnote}{\arabic{footnote}}

\vspace{1.5cm}

{\it Instituto de F\'{\i}sica Te\'orica UAM/CSIC\\
C/ Nicol\'as Cabrera, 13--15,  C.U.~Cantoblanco, E-28049 Madrid, Spain}\\

\vspace{2.5cm}


{\bf Abstract}

\end{center}

\begin{quotation}
  We show that, in all metric theories of gravity with a general covariant
  action, gravity couples to the gravitational energy-momentum tensor in the
  same way it couples to the matter energy-momentum tensor order by order in
  the weak field approximation around flat spacetime. We discuss the relation
  of this property to the Strong Equivalence Principle. We also study the
  gauge transformation properties of the gravitational energy-momentum tensor.
\end{quotation}

\newpage
\pagestyle{plain}
\section{Introduction}

Although General Relativity has passed many experimental tests so far, and in
spite of the general problems of physical theories of higher order in
derivatives,\footnote{Most of these problems are related to the linear
  instability discovered by Ostrogradski in Ref.~\cite{kn:Ostrogradski}
  \cite{Woodard:2015zca} and pedagogically reviewed with its implications in
  Ref.~\cite{Woodard:2015zca}. Nevertheless, see also
  Ref.~\cite{Pavsic:2013noa,Pavsic:2013mja}.} there are many reasons to
consider possible corrections to the Einstein-Hilbert action constructed from
invariants of higher order in the Riemann curvature tensor:\footnote{We are
  only going to consider metric theories based on a general-covariant action
  principle.}

\begin{enumerate}
\item These are the simplest modifications to General Relativity since they do
  not require the explicit introduction of other ``gravitational fields''
  (such as scalar fields). The modified gravity theories remain dynamical
  metric theories, automatically satisfying the Weak Equivalence Principle
  (WEP) if matter only couples minimally to the metric in the
  action.\footnote{Field redefinitions of the metric involving the curvature
    can change this property. We will, henceforth ignore the possibility of
    making this kind of field redefinitions and we will assume our metric is
    the one matter would couple minimally to in the action.}  General
  covariance ensures the on-shell covariant divergenceless of the matter
  energy-momentum tensor.

\item Most theories of Quantum Gravity (in particular Superstring Theories)
  predict an infinite number of higher-order corrections to the
  Einstein-Hilbert action and the effective gravitational theories are
  higher-order gravity theories, but only the lowest order corrections are
  explicitly known. In this context, some of the illnesses of the theories
  with terms of higher order in the curvature can be understood as the result
  of the truncation of a consistent theory.  For Superstring Theories, see,
  \textit{e.g.}~Refs.~\cite{Gross:1986iv,Grisaru:1986vi,Grisaru:1986dk,Grisaru:1986kw,Gross:1986mw,Bergshoeff:1989de,Alvarez-Gaume:2015rwa}

\item The AdS/CFT duals of these higher-order gravity theories are much more
  general than those dual to Einstein-Hilbert gravity and have been proven to
  be very useful in this context. Some interesting examples can be found in,
  \textit{e.g.},
  Refs.~\cite{Brigante:2007nu,Buchel:2009tt,Camanho:2009hu,Cai:2009zv,Buchel:2009sk,Myers:2010jv,Myers:2010xs,Mezei:2014zla,Bueno:2015rda}.

\item Typically, the higher-order terms modify the behaviour of the
  gravitational field in regions of strong curvature. Thus, these corrections
  may play important r\^oles in inflationary cosmology, in black-hole physics
  (see,
  \textit{e.g.}~Refs.~\cite{Boulware:1985wk,Wheeler:1985qd,Behrndt:1998eq,Cai:2001dz,Oliva:2010eb,Myers:2010ru,Lu:2015cqa,Kehagias:2015ata,Hendi:2015hba,Hennigar:2016gkm,Bueno:2016lrh,Hennigar:2017ego,Bueno:2017sui})
  and in the study of spacetime singularities.

\end{enumerate}

In this note we are going to call the theories of gravity resulting from the
addition of terms of higher order in the Riemann curvature tensor
$\mathcal{R}$ or its covariant derivatives
$\nabla\mathcal{R},\nabla^{2}\mathcal{R},\ldots$ invariant under general
coordinate transformations to the Einstein-Hilbert action, that is, theories
with actions of the form

\begin{equation}
S[g] = \frac{1}{\chi^{2}}\int d^{d}x\sqrt{|g|} 
\left\{ R + F(g,\mathcal{R},\nabla\mathcal{R},\cdots) \right\}\, ,   
\hspace{1cm}
\chi^{2}= 16\pi G\, ,
\end{equation}

\noindent
where $R$ is the Ricci scalar as \textit{higher-order
  gravities}.\footnote{Observe that \textit{purely higher-order gravities}
  (without Einstein-Hilbert term) and Palatini ($f(R)$ or else) theories are
  excluded from our considerations. For a review of the latter see,
  \textit{e.g.}~Ref.~\cite{Olmo:2011uz}.}

One of the main obstacles in the study of higher-order gravities is the sheer
number of different combinations of curvature invariants that can be
constructed as the order in curvature and the dimension grow. For this reason,
most of the research has focused on quadratic theories (see,
\textit{e.g.}~Refs.~\cite{Stelle:1976gc,Stelle:1977ry,Julve:1978xn,Boulware:1983td,David:1984uv,Horowitz:1984wv,Deser:2007vs,Alvarez-Gaume:2015rwa})
and in the search for theories with special properties
\cite{kn:Love1970,Lovelock:1971yv,Bergshoeff:2009hq,Myers:2010ru,Oliva:2010eb,Lu:2011zk,Karasu:2016ifk,Bueno:2016xff}. Typically,
one looks for theories which only propagate, in maximally symmetric vacua, a
massless, transverse, graviton or, at the very least, theories that do not
propagate the ghost-like spin-2 mode which generically appears in higher-order
gravities together with scalar modes
\cite{Sisman:2011gz,Tekin:2016vli,Bueno:2016ypa}.  Actually, some of the
higher-order gravities (the $f(R)$ \cite{Sotiriou:2008rp,DeFelice:2010aj,Nojiri:2010wj} or
$f(\mathrm{Lovelock})$ \cite{Sarkar:2013swa,Bueno:2016dol} theories, for
example) can be reformulated as scalar-tensor gravities. In these alternative
representations there are issues related to the use of the Jordan or Einstein
frames and to the qualification of the scalar-mediated interactions, which are
usually different for different kinds of matter, as gravitational or
non-gravitational.

Some of these issues can be avoided by sticking to the original, higher-order,
metric representation. However, if the theory does propagate other modes
besides the massless graviton, one can expect to see some effects associated
to them in that representation as well.\footnote{The existence of other
  representations for a generic higher-order gravity is, by no means,
  guaranteed.}  One of such effects would be a violation of the Strong
Equivalence Principle (SEP) as shown in Ref.~\cite{kn:BG}. The following
paragraph in Section~3.1.2 of Ref.~\cite{Will:2014kxa} summarizes very well
our current understanding: after arguing that only metric theories of gravity
have a chance of satisfying the SEP it is stated that

\begin{quotation}
  \textit{Empirically it has been found that almost every metric theory other
    than GR introduces auxiliary gravitational fields, either dynamical or
    prior geometric, and thus predicts violations of SEP at some level (here
    we ignore quantum-theory inspired modifications to GR involving
    ``$R^{2}$'' terms).}
\end{quotation}

The reasons why \textit{quantum-theory inspired modifications to GR involving
  ``$R^{2}$'' terms}, which are precisely the subject of this note, should be
kept out of the discussion are not clear to us.  As we have just mentioned,
some of them admit a scalar-tensor representation which suggests violations of
SEP. On the other hand, there are higher-order theories such as the
\textit{Einsteinian cubic gravity} of Ref.~\cite{Bueno:2016xff} which only
propagate a transverse, traceless, massless spin-2 mode. Do they violate the
SEP? Is it possible to make a general statement concerning the SEP in
higher-order gravities?

The difference between the Einstein and the Strong Equivalence Principles (EEP
and SEP, resp.)\footnote{A general formulation of the different Equivalence
  Principles can be found in, \textit{e.g.}~\cite{DiCasola:2013iia}.} is that
the second extends the former to situations in which the presence of
gravitational fields is significant. The main effects of gravitational fields
in a local system in free fall are

\begin{enumerate}
\item Contributions to the binding energy of macroscopic bodies or Keplerian
  systems.

\item Determination, through the curvature, of the boundary conditions for
  locally Minkowskian metrics.
\end{enumerate}

The second effect, which falls out of the scope of this note, is responsible
for most of the violations of the SEP in higher-order gravities since, via the
curvature of a external gravitational field, free-falling systems can suffer
spacetime position-dependent effects. Often they are associated to the
emergence of an effective spacetime position-dependent Newton constant,
although this concept is far from having a unique and clear definition.

The consequences of the first effect depend on the nature of the coupling of
gravity to gravity: the SEP states that it must be identical to the coupling
of gravity to other forms of matter/energy so that self-gravitation does not
modify the free fall motion of massive
bodies.\footnote{\label{foot:selfcoupling} The necessity of the self-coupling
  of the relativistic gravitational field arises naturally in the construction
  of a classical interacting special-relativistic theory of gravity, as
  reviewed in Ref.~\cite{Ortin:2015hya}. The Lorentz invariance of the S
  matrix of the quantized theory requires the graviton field to couple to the
  total energy-momentum tensor \cite{Weinberg:1964ew,kn:We3}, which , in the
  long wavelength limit must be that derived from General Relativity in the
  weak-field limit \cite{Boulware:1974sr}. At the same time, the consistency
  of the self-coupling requires the introduction of an infinite number of
  corrections as first noticed by Gupta \cite{Gupta:1954zz,kn:Gup2} who
  devised the possible construction of the full theory by demanding
  consistency of the self-coupling of the gravitational field at all orders
  (the so-called ``Gupta program''). (Obviously, the meaning of
  \textit{consistency} is one of the keys in this problem.)  It has been
  argued that the resulting theory is equivalent to General Relativity
  \cite{Kraichnan:1955zz,Deser:1969wk,Butcher:2009ta,Deser:2009fq} although
  this conclusion seems to depend strongly on the starting point, as shown in
  Ref.~\cite{Barcelo:2014mua} where a different theory, equivalent to
  unimodular gravity, was found. Although the theories that we are considering
  have an Einstein-Hilbert term and, therefore, a Fierz-Pauli term in the
  weak-field limit (which seems to lead unavoidably to General Relativity) we
  are not going to use explicitly this fact in what follows and our results
  will be valid for more general classes of general covariant
  theories.}${}^{'}$\footnote{The violation of this part of the SEP would give
  rise to the Nordtvedt effect \cite{Nordtvedt:1968qr,Nordtvedt:1968qs}. 
  Section~\ref{sec-EMtensors}.
} In metric theories, thus, the SEP demands that
the metric couples to the gravitational energy-momentum tensor in the same way
it couples to the energy-momentum tensor of any other kind of matter. This
property only makes sense in the weak-field limit in which a
special-relativistic gravitational energy-momentum tensor can be
(non-uniquely) defined, but it is a property that any metric theory of gravity
can be tested for unambiguously. This is what we are going to do for
higher-order gravities.

Let us start by formulating more precisely this property.

If we expand the gravitational field around the Minkowski spacetime in the
weak field approximation $g_{\mu\nu}=\eta_{\mu\nu}+\chi h_{\mu\nu}$ then the
SEP requires the equations of motion of the gravitational field to second
order in $h_{\mu\nu}$ to have the form\footnote{Here we follow the notation
  and conventions of Ref.~\cite{Ortin:2015hya}. The consequences of having
  different couplings for the gravitational and matter energy-momentum tensors
  ($\chi^{\prime}$ and $\chi$) at the linear level were first studied by
  Kraichnan in Ref.~\cite{kn:Kra2}.}

\begin{equation}
\label{eq:eom}
\mathcal{D}^{(0)\, \mu\nu} = \chi (t^{(0)\, \mu\nu}+T_{\rm matter}^{\mu\nu})\, ,  
\end{equation}

\noindent
where $\mathcal{D}^{(0)\, \mu\nu}$ is a wave operator that acts linearly on
$h_{\mu\nu}$ but which is of arbitrary order in derivatives, computed from the
quadratic (zeroth-order) term in the action, $t^{(0)\, \mu\nu}$ is
zeroth-order gravitational energy-momentum tensor in Minkowski spacetime
computed from the same term in the action, quadratic in $h_{\mu\nu}$ but also
of arbitrary order in derivatives and $T_{\rm matter}^{\mu\nu}$ is the matter
energy-momentum tensor in Minkowski spacetime computed according to
Rosenfeld's prescription \cite{kn:Ros}.\footnote{That is: coupling matter
  minimally to the metric $g_{\mu\nu}$ and computing
\begin{equation}
T_{\rm matter}^{\mu\nu}=
\left.
2 
\frac{\delta S_{\rm matter}}{\delta g_{\mu\nu}} 
\right|_{g_{\mu\nu}=\eta_{\mu\nu}}.
\end{equation}
In the context of the weak-field expansion this is the energy-momentum tensor
that occurs naturally coupled to $h_{\mu\nu}$.
}

The consistency of the above equation will be the main ingredient of our
discussion. Observe that, if it can be derived from the lowest-order terms of an
action\footnote{We ignore the matter fields from now onwards.}

\begin{equation}
-2\frac{\delta S}{\delta h_{\mu\nu}} 
= 
\mathcal{D}^{(0)\, \mu\nu} -\chi t^{(0)\, \mu\nu} +\ldots\, ,  
\hspace{1cm}
S
= 
S^{(0)}+\chi S^{(1)} +\chi^{2} S^{(2)}+\ldots 
\end{equation}

\noindent

\noindent
as we have assumed, the wave operator originates in the variation of the 
zeroth-order term

\begin{equation}
\mathcal{D}^{(0)\, \mu\nu}  
=
-2\frac{\delta S^{(0)}}{\delta h_{\mu\nu}}\, ,
\end{equation}

\noindent
while the gravitational energy-momentum tensor must originate in the
first-order term in the action

\begin{equation}
\label{eq:toprove}
t^{(0)\, \mu\nu}
=
2\frac{\delta S^{(1)}}{\delta h_{\mu\nu}}\, .
\end{equation}

The variation of $S^{(1)}$ will always give some term quadratic in
$h_{\mu\nu}$, but the SEP demands that this term is precisely the
energy-momentum tensor of the gravitational field corresponding to $S^{(0)}$.

The rest of this note is devoted to proving that this property, which requires
a very precise relation between $S^{(0)}$ and $S^{(1)}$, holds in all
higher-order gravities. We will start by recalling some well- and
less-well-known facts about special-relativistic energy-momentum tensors of
theories of higher-order in derivatives in Section~\ref{sec-EMtensors}. In
Section~\ref{sec-perturbativegravity} we will study the implications of
general covariance concerning the gravitational energy-momentum tensor in the
context of perturbative gravity and in Section~\ref{sec-EmtensorHOgravity} we
will use them to prove Eq.~(\ref{eq:toprove}). Section~\ref{sec-discussion}
contains our conclusions and a discussion of how the SEP may be satisfied when
terms of higher order in $h_{\mu\nu}$ are considered.

\section{Energy-momentum tensors}
\label{sec-EMtensors}

Let us consider a $d$-dimensional field theory with action

\begin{equation}
S = \int d^{d}x\, \mathcal{L}\, ,  
\end{equation}

\noindent
where the Lagrangian $\mathcal{L}$ is a function of the field $\phi$ and its
derivatives $\partial_{\mu}\phi$, $\partial_{\mu}\partial_{\nu}\phi$,
$\partial_{\mu}\partial_{\nu}\partial_{\rho}\phi,\ldots$ up to an arbitrary
order.

Imposing adequate conditions for the boundary values of the field and a number
of its derivatives, demanding the extremization of the above action leads to
the Euler-Lagrange equations

\begin{equation}
\frac{\delta S}{\delta \phi} 
=
\frac{\partial \mathcal{L}}{\partial\phi}  
-\partial_{\mu}\frac{\partial \mathcal{L}}{\partial\partial_{\mu}\phi}
+\partial_{\mu}\partial_{\nu}
\frac{\partial \mathcal{L}}{\partial\partial_{\mu}\partial_{\nu}\phi}
-\partial_{\mu}\partial_{\nu}\partial_{\rho}
\frac{\partial \mathcal{L}}{\partial\partial_{\mu}\partial_{\nu}\partial_{\rho}\phi}
+\ldots
\end{equation}

According to the first Noether theorem, the invariance of the above action
under constant displacements of the coordinates $\delta x^{\mu}$, with $\delta
\phi= \delta x^{\mu}\partial_{\mu}\phi$ leads to the relation

\begin{equation}
\label{eq:mainpropertygeneral}
\partial_{\mu}t_{\rm can}{}^{\mu}{}_{\nu} 
=
\frac{\delta S}{\delta \phi}\partial_{\nu}\phi\, , 
\end{equation}

\noindent
where 

\begin{equation}
t_{\rm can}{}^{\mu}{}_{\nu}
\equiv
\eta^{\mu}{}_{\nu}\mathcal{L}
-\frac{\partial \mathcal{L}}{\partial\partial_{\mu}\phi}\partial_{\nu}\phi  
-\frac{\partial \mathcal{L}}{\partial\partial_{\mu}\partial_{\alpha}\phi}
\partial_{\nu}\partial_{\alpha}\phi  
+\partial_{\alpha}
\left(
\frac{\partial \mathcal{L}}{\partial\partial_{\mu}\partial_{\alpha}\phi}
\right)
\partial_{\nu}\phi  
+\ldots
\end{equation}

\noindent
is the canonical energy-momentum tensor.  The above relation implies its
on-shell conservation (\textit{i.e.}~when $\delta S/\delta \phi=0$), but we
will need the complete off-shell relation, which, for the gravitational field,
at lowest order, takes the form 

\begin{equation}
\label{eq:mainproperty}
\partial_{\mu}t^{(0)}_{\rm can}{}^{\mu}{}_{\nu} 
=
\frac{\delta S^{(0)}}{\delta h_{\rho\sigma}}\partial_{\nu}h_{\rho\sigma}\, . 
\end{equation}

Noether currents and, in particular, the canonical energy-momentum tensor, are
not unambiguously defined: one can add to them terms proportional to the
equations of motion, that will vanish on shell, and superpotential terms of
the form 

\begin{equation}
\partial_{\rho}\Psi^{\rho\mu}{}_{\nu}\, ,
\,\,\,\,\,
\mbox{with}
\,\,\,\,\,
\Psi^{\rho\mu}{}_{\nu}
=
\Psi^{[\rho\mu]}{}_{\nu}\, .
\end{equation}

\noindent
The positive side of this ambiguity is that it can be used to make
gauge-invariant or symmetrize the canonical energy-momentum tensor. Belinfante
\cite{kn:Bel} found a systematic procedure to symmetrize the canonical
energy-momentum tensor of fundamental fields that gives the same result as the
Rosenfeld prescription for many fields, but not for the gravitational field
$h_{\mu\nu}$ unless one adds terms proportional to the equations of motion
\cite{Ortin:2015hya}. In the General Relativity case, it was shown in
Refs.~\cite{Feynman:1996kb,Ortin:2015hya} that the gravitational
energy-momentum tensor which is singled-out by the theory (and satisfies the
consistency condition Eq.~(\ref{eq:toprove})) is completely determined by
gauge invariance and it is related to the canonical one by a superpotential
and on-shell-vanishing terms. It was also shown that using other
energy-momentum tensors (as done by Thirring in Ref.~\cite{kn:Thir2}, for
instance) leads to the wrong value for the secular shift of Mercury's
perihelion. The same principle should determine the gravitational
energy-momentum tensor in higher-order gravities.\footnote{It should also
  determine the higher-order corrections for Fierz-Pauli that eventually lead
  to General Relativity. In the search for theories with consistent
  self-coupling of the gravitational field $h_{\mu\nu}$ (Gupta's program)
  mentioned in footnote~\ref{foot:selfcoupling}, if one just tries to couple
  the gravitational field to its own energy-momentum tensor, the ambiguity in
  the definition of the latter turns the problem into the problem of which is
  the ``right'' energy-momentum tensor. Even if one can show that there is a
  prescription that in the end gives General Relativity, as in
  Ref.~\cite{Butcher:2009ta}, one needs some physical principle to justify it
  and its uniqueness. Gauge invariance plays here this r\^ole.} On the other
hand, given a conserved 2-index tensor, quadratic in the gravitational field,
we will identify it with the energy-momentum tensor if it differs from the
canonical one by superpotential and on-shell-vanishing terms.

Let us know examine in more details how gauge invariance acts in this context.

\section{Perturbative gravity and gauge invariance}
\label{sec-perturbativegravity}

Higher-order gravities are covariant under general coordinate transformations
$\delta_{\xi}S=0$. In the weak-field limit this symmetry manifests itself
order by order in $\chi$

\begin{equation}
\delta_{\xi}   
=
\delta^{(0)}_{\xi}+\chi \delta_{\xi}^{(1)} +\chi^{2} \delta_{\xi}^{(2)}+\ldots 
\end{equation}

The variation of the gravitational fields $h_{\mu\nu}$, only has zeroth- and
first-order terms: zeroth- and first-order transformations

\begin{equation}
\label{eq:gaugetrans1}
\delta_{\xi}h_{\mu\nu}   
=
\left(\delta^{(0)}_{\xi}+\chi \delta_{\xi}^{(1)}\right)h_{\mu\nu}\, ,
\end{equation}

\noindent
with

\begin{equation}
\label{eq:gaugetrans2}
\delta^{(0)}_{\xi}h_{\mu\nu} = 2\partial_{(\mu}\xi_{\nu)}\, ,
\hspace{1cm}
\delta^{(1)}_{\xi}h_{\mu\nu} = \pounds_{\xi}h_{\mu\nu} 
= 
\xi^{\rho}\partial_{\rho}h_{\mu\nu}+2\partial_{(\mu}\xi^{\rho}h_{\nu)\rho}\, .
\end{equation}

\noindent
The parameter $\xi^{\mu}$ is completely arbitrary and it does not satisfy any
constraints restricting the general covariance of the original theory. This
condition excludes the unimodular theories which are invariant under the above
transformations for divergenceless parameters $\partial_{\mu}\xi^{\mu}=0$
only \cite{Kreuzer:1989ec}.

These transformations relate terms of consecutive orders in the action:

\begin{eqnarray}
\delta^{(0)}_{\xi}S^{(0)}  & = & 0\, ,
\\
& & \nonumber \\
\delta^{(0)}_{\xi}S^{(n)}+\delta^{(1)}_{\xi}S^{(n-1)} & = & 0\, ,
\,\,\,\,\,
\mbox{for}
\,\,\,\,\,
n\leq 1\, . 
\end{eqnarray}

To our purposes, each of these relations can be seen as an independent gauge
symmetry, and there is a Noether (also called gauge, or Bianchi) identity
associated to each of these local symmetries via Noether's second
theorem. Using the explicit form of the transformations of the gravitational
field one arrives, after integration by parts and elimination of total
derivatives, to the first two Noether
identities\footnote{\label{foot:unimodular} The restricted gauge invariance of
  linearized unimodular gravity leads to different Noether identities
  \cite{Alvarez:2005iy,Alvarez:2006uu}. The classical equations of motion of
  unimodular gravity \cite{kn:Eins} are those of General Relativity with a
  cosmological constant and, presumably, this theory must enjoy the same
  property we are proving here, although it must arise in a more complicated
  way.}


\begin{eqnarray}
\label{eq:Noetheridentity0}
\partial_{\mu}\frac{\delta S^{(0)}}{\delta h_{\mu\nu}}
& = &
0\, ,
\\
& & \nonumber \\
\label{eq:Noetheridentity}
\partial_{\mu}\frac{\delta S^{(1)}}{\delta h_{\mu\nu}}
& = & 
-\partial_{\mu} \left(\frac{\delta S^{(0)}}{\delta h_{\mu\rho}}h_{\rho}{}^{\nu} \right)
+\tfrac{1}{2}\frac{\delta S^{(0)}}{\delta
  h_{\rho\sigma}}\partial^{\nu}h_{\rho\sigma}\, .
\end{eqnarray}

Together, if Eq.~(\ref{eq:toprove}) is satisfied, they are consistent with the
on-shell conservation of the gravitational energy-momentum tensor to lowest
order in $\chi$: if we take the divergence of both sides of the equation of
motion (\ref{eq:eom}) and use the first identity, we get

\begin{equation}
\chi \partial_{\mu}t^{(0)\, \mu\nu}=0\, .  
\end{equation}

\noindent
The second identity says that

\begin{equation}
\chi \partial_{\mu}t^{(0)\, \mu\nu}
=
-2\chi\partial_{\mu} \left(\frac{\delta S^{(0)}}{\delta h_{\mu\rho}}h_{\rho\nu} \right)
+\chi\frac{\delta S^{(0)}}{\delta h_{\rho\sigma}}\partial_{\nu}h_{\rho\sigma}\, ,
\end{equation}

\noindent
and, using again the equations of motion

\begin{equation}
\chi \partial_{\mu}t^{(0)\, \mu\nu}
=
\mathcal{O}(\chi^{2})\, ,
\,\,\,\,\,
\mbox{on shell.}
\end{equation}

Apart from this consistency check, let us stress that the above identities
Eqs.~(\ref{eq:Noetheridentity0}),(\ref{eq:Noetheridentity}) hold off-shell for
any higher-order gravity.

\section{The energy-momentum tensor of higher-order gravity}
\label{sec-EmtensorHOgravity}

We are now ready to prove that Eq.~(\ref{eq:toprove}) holds; that is: that
$2\delta S^{(1)}/\delta h_{\mu\nu}$ can be identified with one of the
special-relativistic energy-momentum tensors that one can associate to
$S^{(0)}$, such as the canonical energy-momentum tensor $t^{(0)}_{\rm
  can}{}^{\mu}{}_{\nu}$ plus some on-shell-vanishing terms (to lowest order in
$\chi$!) and a superpotential term.

Using in the Noether identity Eq.~(\ref{eq:Noetheridentity}) the property
Eq.~(\ref{eq:mainproperty}) we can rewrite it as follows:

\begin{equation}
\partial_{\mu}\left(2 \frac{\delta S^{(1)}}{\delta h_{\mu\nu}}\right)
=
\partial_{\mu} \left(
t^{(0)}_{\rm can}{}^{\mu\nu}
-2\frac{\delta S^{(0)}}{\delta h_{\mu\rho}}h_{\rho}{}^{\nu}
 \right)\, ,
\end{equation}

\noindent
from which it follows that 

\begin{equation}
2 \frac{\delta S^{(1)}}{\delta h_{\mu\nu}}
=
t^{(0)}_{\rm can}{}^{\mu\nu}
-2\frac{\delta S^{(0)}}{\delta h_{\mu\rho}}h_{\rho}{}^{\nu}
+\partial_{\rho}\Psi^{(0)\,\rho\mu\nu}\, ,
\,\,\,\,\,
\mbox{with}
\,\,\,\,\,
\Psi^{(0)\, \rho\mu\nu}
=
-\Psi^{(0)\,\mu\rho\nu}\, .
\end{equation}

Given that the second term in the r.h.s.~vanishes on-shell at the order in
$\chi$ we are working at,\footnote{Equivalently, if we use the equations of
  motion, that term would be of $\mathcal{O}(h^{3})$.} the l.h.s.~of the above
equation can be identified with a special-relativistic gravitational
energy-momentum tensor associated to the zeroth-order action $S^{(0)}$,
proving Eq.~(\ref{eq:toprove}).

\section{Extension to higher orders}
\label{sec-higherorders}

Can we extend this $\mathcal{O}(h)$ result to higher orders? 

The $n$th Noether identity is

\begin{equation}
\label{eq:Noetheridentityn}
\partial_{\mu}\frac{\delta S^{(n)}}{\delta h_{\mu\nu}}
=
-\partial_{\mu} \left(\frac{\delta S^{(n-1)}}{\delta h_{\mu\rho}}h_{\rho}{}^{\nu} \right)
+\tfrac{1}{2}\frac{\delta S^{(n-1)}}{\delta
  h_{\rho\sigma}}\partial^{\nu}h_{\rho\sigma}\, ,
\end{equation}

\noindent
and using again Eq.~(\ref{eq:mainproperty}) and the same reasoning as in the
$n=1$ case, we arrive to

\begin{equation}
2 \frac{\delta S^{(n)}}{\delta h_{\mu\nu}}
=
t^{(n-1)}_{\rm can}{}^{\mu\nu}
-2\frac{\delta S^{(n-1)}}{\delta h_{\mu\rho}}h_{\rho}{}^{\nu}
+\partial_{\rho}\Psi^{(n-1)\,\rho\mu\nu}\, ,
\,\,\,\,\,
\mbox{with}
\,\,\,\,\,
\Psi^{(n-1)\, \rho\mu\nu}
=
-\Psi^{(n-1)\,\mu\rho\nu}\, .
\end{equation}

\noindent
where $t^{(n-1)}_{\rm can}{}^{\mu\nu}$ is the contribution to the
gravitational special-relativistic canonical energy-momentum tensor coming
from the $S^{(n-1)}$. Adding all these relations to order $n$ multiplied by
the corresponding power of $\chi$ we get

\begin{eqnarray}
2 
\left( 
\chi\frac{\delta S^{(1)}}{\delta h_{\mu\nu}}
+\chi^{2}\frac{\delta S^{(2)}}{\delta h_{\mu\nu}}
+\cdots +
\chi^{n}\frac{\delta S^{(n)}}{\delta h_{\mu\nu}}
\right)\,\, =
\hspace{-4cm}
& & \nonumber \\
& & \nonumber \\
& & 
\chi t^{(0)}_{\rm can}{}^{\mu\nu}
+\chi^{2}t^{(1)}_{\rm can}{}^{\mu\nu}
+\cdots +
\chi^{n}t^{(n-1)}_{\rm can}{}^{\mu\nu}
\nonumber \\
& & \nonumber \\
& & 
-2\chi
\left( 
\frac{\delta S^{(0)}}{\delta h_{\mu\nu}}
+\chi\frac{\delta S^{(1)}}{\delta h_{\mu\nu}}
+\cdots +
\chi^{n-1}\frac{\delta S^{(n-1)}}{\delta h_{\mu\nu}}
\right)
h_{\rho}{}^{\nu}
\nonumber \\
& & \nonumber \\
& & 
+\partial_{\rho}
\left(
\chi\Psi^{(0)\,\rho\mu\nu}
+\chi^{2}\Psi^{(1)\,\rho\mu\nu}
+\cdots +
+\chi^{n}\Psi^{(n-1)\,\rho\mu\nu}
\right)
\, .
\label{eq:esaeq}
\end{eqnarray}

The $n$th-order equations of motion say that 

\begin{equation}
\frac{\delta S^{(0)}}{\delta h_{\mu\nu}}
+\chi\frac{\delta S^{(1)}}{\delta h_{\mu\nu}}
+\cdots +
\chi^{n-1}\frac{\delta S^{(n-1)}}{\delta h_{\mu\nu}}
=
-\chi^{n}\frac{\delta S^{(n)}}{\delta h_{\mu\nu}}  
\end{equation}

\noindent
and, therefore, the term in the second line of the r.h.s.~of
Eq.~(\ref{eq:esaeq}) vanishes up to this order on-shell and the term in the
l.h.s.~can be identified with a special-relativistic energy-momentum tensor
for the gravitational field.

\section{Gauge transformations of the energy-momentum tensor}
\label{sec-emtransformations}

The lowest-order energy-momentum tensor of the spin-2 field, $t^{(0)}_{{\rm
    GR}\, \mu\nu}$ is not expected to be invariant under the
$\delta_{\xi}^{(0)}$ gauge transformations. Its gauge transformation rule must
be completely determined by the gauge-invariance of the action and the same
must be true for $t^{(0)}_{\mu\nu}$ and its higher order in $h$ corrections in
higher-order theories. In order to find this gauge transformation rule we
proceed as follows: we first compute the gauge trasformations
Eqs.~(\ref{eq:gaugetrans1}) and (\ref{eq:gaugetrans2}) of the action $\delta
_{\xi}S[h]$ order by order in $\chi$. By assumption $\delta _{\xi}S[h]=0$ up
to total erivatives.  Then we take another general variation with respect to
$h_{\mu\nu}$, $\delta\delta _{\xi}S[h]=0$, taking into account that
$\frac{\delta\delta^{(0)}_{\xi}h_{\mu\nu}}{\delta h_{\alpha\beta}}=0$ but
$\frac{\delta\delta^{(1)}_{\xi}h_{\mu\nu}}{\delta h_{\alpha\beta}}\neq 0$, so
that, integrating by parts

\begin{equation}
 \int d^{d}x
\frac{\delta S^{(n)}}{\delta h_{\mu\nu}}\frac{\delta\delta^{(1)}_{\xi}h_{\mu\nu}}{\delta h_{\alpha\beta}}
=
 \int d^{d}x
\left\{
-\partial_{\rho}\left(\xi^{\rho} 
\frac{\delta S^{(n)}}{\delta h_{\alpha\beta}}\right)
+2\partial^{(\alpha}\xi_{\rho} 
\frac{\delta S^{(n)}}{\delta h_{\beta)\rho}}
\right\}\, .
\end{equation}

\noindent
Next, we interchange the variations of the $S^{(n)}$, arriving to

\begin{eqnarray}
\delta \delta_{\xi}S[h]
& = &
 \int d^{d}x
\left\{
\delta^{(0)}_{\xi}\frac{\delta S^{(0)}}{\delta h_{\alpha\beta}}
+
\chi\left[
\delta^{(1)}_{\xi}\frac{\delta S^{(0)}}{\delta h_{\alpha\beta}}
+
\delta^{(0)}_{\xi}\frac{\delta S^{(1)}}{\delta h_{\alpha\beta}}
-\partial_{\rho}\left(\xi^{\rho} 
\frac{\delta S^{(0)}}{\delta h_{\alpha\beta}}\right)
\right.
\right.
\nonumber \\
& & \nonumber \\
& & 
\left.
\left.
+2\partial^{(\alpha}\xi_{\rho} 
\frac{\delta S^{(0)}}{\delta h_{\beta)\rho}}
\right]
+\cdots
\right\}\delta h_{\alpha\beta}
=0\, .
\end{eqnarray}
 
Finally, the standard arguments lead to the identities

\begin{equation}
\delta^{(0)}_{\xi}\frac{\delta S^{(n)}}{\delta h_{\alpha\beta}}
=
-\delta^{(1)}_{\xi}\frac{\delta S^{(n-1)}}{\delta h_{\alpha\beta}}
+\partial_{\rho}\left(\xi^{\rho} 
\frac{\delta S^{(n-1)}}{\delta h_{\alpha\beta}}\right)
-2\partial^{(\alpha}\xi_{\rho} 
\frac{\delta S^{(n-1)}}{\delta h_{\beta)\rho}}\, ,
\,\,\,\,\,
\forall n\geq 0\, .
\end{equation}

\noindent
Obviously, the sums of the terms $n=1,\cdots N$ and those of the terms
$n=0,\cdots N-1$ satisfy the same relation.

For GR at lowest order, these identities imply 

\begin{eqnarray}
\delta^{(0)}_{\xi} \mathcal{D}^{(0)\, \mu\nu}
& = &
0\, ,
\\
& & \nonumber \\
\delta^{(0)}_{\xi} t^{(0)\, \mu\nu}_{\rm GR} 
& = &
\delta^{(1)}_{\xi} \mathcal{D}^{(0)\, \mu\nu}
+\partial_{\rho}\left(\xi^{\rho} \mathcal{D}^{(0)\, \mu\nu}\right)
-2\partial^{(\mu|}\xi_{\rho} \mathcal{D}^{(0)\, |\nu)\rho}\, ,
\end{eqnarray}

\noindent
which can be checked to hold using the Fierz-Pauli equation of motion and the
explicit expression for $ t^{(0)\, \mu\nu}_{\rm GR}$ in Eq.~(3.200) of
Ref.~\cite{Ortin:2015hya}. The equation 

\begin{equation}
\mathcal{D}^{(0)\, \mu\nu} = \chi t^{(0)\, \mu\nu}_{\rm GR}\, ,
\end{equation}

\noindent
is only invariant under
$(\delta^{(0)}_{\xi}+\chi\delta^{(1)}_{\xi})h_{\mu\nu}$ and only 
up to terms proportional to $\mathcal{D}^{(0)\, \mu\nu}$ and its derivatives
which, on-shell, are of order $\mathcal{O}(\chi^{2})$.

\section{Discussion}
\label{sec-discussion}

In this note we have shown how general covariance in metric theories leads to
equations of motion for the perturbative gravitational field sourced by the
gravitational energy-momentum tensor computed to any order. 

As we have stressed in the introduction, this result shows that in all the
theories under consideration the gravitational binding energy of massive
bodies will not modify their free-fall motion, which is one of the
requirements of the SEP. Violations of other requirements of the SEP are
generally expected and our result has nothing to say about them. It can be
said that we have only shown a ``microscopic version of the SEP.''

This result has nothing to say about the correctness of the theories,
either. It only shows that they are internally consistent and that this is due
to the gauge invariance that follows from general covariance.

Our result also raises the possibility of carrying out Gupta's program
starting with linearized theories of higher order in derivatives at the evel
of the actin or at the level of the equations of motion
(Section~\ref{sec-emtransformations}). Each correction is strongly
constrained by the requirement of gauge invariance, suggesting that there is a
unique correction at each order and that, the whole theory is completely
determined by the lowest order that describes the free one.

As discussed in footnote~\ref{foot:unimodular}, it would also be very
interesting to understand if and how the unimodular theory of gravity (and its
higher-order generalizations) enjoy similar properties to understand to which
extent general covariance is fundamental. Other extensions of this work that
would interesting to study are the analysis in De Sitter and anti-De Sitter
backgrounds for which selfconsitency is known to lead to GR plus a
cosmological constant \cite{Deser:1987uk}.

\section*{Acknowledgments}

The author would like to thank E.~\'Alvarez, F.~Barbero, P.~Bueno, P.A.~Cano,
R.~Carballo-Rubio, S.~Deser, R.~Emparan, L.J.~Garay, G.J.~Olmo and
J.M.M.~Senovilla for many interesting and useful comments on this topic and on
the draft. He would also like to express a special thanks to the Mainz
Institute for Theoretical Physics (MITP) for its hospitality and support and
M.M.~Fern\'andez for her permanent support.  This work has been supported in
part by the MINECO/FEDER, UE grant FPA2015-66793-P and the Centro de
Excelencia Severo Ochoa Program grant SEV-2012-0249.



\begin{thebibliography}{99}

\bibitem{kn:Ostrogradski}
M.~Ostrogradski, 
Mem. Ac. St. Petersbourg \textbf{VI 4} (1850) 385.

\bibitem{Woodard:2015zca}
R.~P.~Woodard,
``Ostrogradsky's theorem on Hamiltonian instability,''
Scholarpedia {\bf 10} (2015) no.8,  32243.
\doi{10.4249/scholarpedia.32243}
[\arxiv{1506.02210} [hep-th]].

\bibitem{Pavsic:2013noa}
M.~Pav\v{s}i\v{c},
``Stable Self-Interacting Pais-Uhlenbeck Oscillator,''
Mod.\ Phys.\ Lett.\ A {\bf 28} (2013) 1350165.
\doi{10.1142/S0217732313501654}
[\arxiv{1302.5257} [gr-qc]].

\bibitem{Pavsic:2013mja}
M.~Pav\v{s}i\v{c},
``Pais-Uhlenbeck Oscillator with a Benign Friction Force,''
Phys.\ Rev.\ D {\bf 87} (2013) no.10,  107502.
\doi{10.1103/PhysRevD.87.107502}
[\arxiv{1304.1325} [gr-qc]].

\bibitem{Gross:1986iv}
D.~J.~Gross and E.~Witten,
``Superstring Modifications of Einstein's Equations,''
Nucl.\ Phys.\ B {\bf 277}, 1 (1986).
\doi{10.1016/0550-3213(86)90429-3}

\bibitem{Grisaru:1986vi}
M.~T.~Grisaru and D.~Zanon,
``$\sigma$-Model Superstring Corrections to the 
Einstein-Hilbert Action,''
Phys.\ Lett.\ B {\bf 177} (1986) 347.
\doi{10.1016/0370-2693(86)90765-3}

\bibitem{Grisaru:1986dk}
M.~T.~Grisaru, A.~E.~M.~van de Ven and D.~Zanon,
``Two-Dimensional Supersymmetric Sigma Models on Ricci Flat 
K\"ahler Manifolds Are Not Finite,''
Nucl.\ Phys.\ B {\bf 277} (1986) 388.
\doi{10.1016/0550-3213(86)90448-7}

\bibitem{Grisaru:1986kw}
M.~T.~Grisaru, A.~E.~M.~van de Ven and D.~Zanon,
``Four Loop Divergences for the N=1 Supersymmetric Nonlinear 
Sigma Model in Two-Dimensions,''
Nucl.\ Phys.\ B {\bf 277} (1986) 409.
\doi{10.1016/0550-3213(86)90449-9}

\bibitem{Gross:1986mw}
D.~J.~Gross and J.~H.~Sloan,
``The Quartic Effective Action for the Heterotic String,''
Nucl.\ Phys.\ B {\bf 291} (1987) 41.
\doi{10.1016/0550-3213(87)90465-2}

\bibitem{Bergshoeff:1989de}
E.~A.~Bergshoeff and M.~de Roo,
``The Quartic Effective Action of the Heterotic String and 
Supersymmetry,''
Nucl.\ Phys.\ B {\bf 328} (1989) 439.
\doi{10.1016/0550-3213(89)90336-2}

\bibitem{Alvarez-Gaume:2015rwa}
L.~Alvarez-Gaume, A.~Kehagias, C.~Kounnas, D.~L\"ust and A.~Riotto,
``Aspects of Quadratic Gravity,''
Fortsch.\ Phys.\  {\bf 64} (2016) no.2-3,  176.
\doi{10.1002/prop.201500100}
[\arxiv{1505.07657} [hep-th]].

\bibitem{Brigante:2007nu}
M.~Brigante, H.~Liu, R.~C.~Myers, S.~Shenker and S.~Yaida,
``Viscosity Bound Violation in Higher Derivative Gravity,''
Phys.\ Rev.\ D {\bf 77} (2008) 126006.
\doi{10.1103/PhysRevD.77.126006}
[\arxiv{0712.0805} [hep-th]].

\bibitem{Buchel:2009tt}
A.~Buchel and R.~C.~Myers,
``Causality of Holographic Hydrodynamics,''
JHEP {\bf 0908} (2009) 016.
\doi{10.1088/1126-6708/2009/08/016}
[\arxiv{0906.2922} [hep-th]].

\bibitem{Camanho:2009hu}
X.~O.~Camanho and J.~D.~Edelstein,
``Causality in AdS/CFT and Lovelock theory,''
JHEP {\bf 1006} (2010) 099.
\doi{10.1007/JHEP06(2010)099}
[\arxiv{0912.1944} [hep-th]].

\bibitem{Cai:2009zv}
R.~G.~Cai, Z.~Y.~Nie, N.~Ohta and Y.~W.~Sun,
``Shear Viscosity from Gauss-Bonnet Gravity with a Dilaton Coupling,''
Phys.\ Rev.\ D {\bf 79} (2009) 066004.
\doi{10.1103/PhysRevD.79.066004}
[\arxiv{0901.1421} [hep-th]].

\bibitem{Buchel:2009sk}
A.~Buchel, J.~Escobedo, R.~C.~Myers, M.~F.~Paulos, A.~Sinha and M.~Smolkin,
``Holographic GB gravity in arbitrary dimensions,''
JHEP {\bf 1003} (2010) 111.
\doi{10.1007/JHEP03(2010)111}
[\arxiv{0911.4257} [hep-th]].

\bibitem{Myers:2010jv}
R.~C.~Myers, M.~F.~Paulos and A.~Sinha,
``Holographic studies of quasi-topological gravity,''
JHEP {\bf 1008} (2010) 035.
\doi{10.1007/JHEP08(2010)035}
[\arxiv{1004.2055} [hep-th]].

\bibitem{Myers:2010xs}
R.~C.~Myers and A.~Sinha,
``Seeing a c-theorem with holography,''
Phys.\ Rev.\ D {\bf 82} (2010) 046006.
\doi{10.1103/PhysRevD.82.046006}
[\arxiv{1006.1263} [hep-th]].

\bibitem{Mezei:2014zla}
M.~Mezei,
``Entanglement entropy across a deformed sphere,''
Phys.\ Rev.\ D {\bf 91} (2015) no.4,  045038.
\doi{10.1103/PhysRevD.91.045038}
[\arxiv{1411.7011} [hep-th]].

\bibitem{Bueno:2015rda}
P.~Bueno, R.~C.~Myers and W.~Witczak-Krempa,
``Universality of corner entanglement in conformal field theories,''
Phys.\ Rev.\ Lett.\  {\bf 115} (2015) 021602.
\doi{10.1103/PhysRevLett.115.021602}
[\arxiv{1505.04804} [hep-th]].

\bibitem{Boulware:1985wk}
D.~G.~Boulware and S.~Deser,
``String Generated Gravity Models,''
Phys.\ Rev.\ Lett.\  {\bf 55} (1985) 2656.
\doi{10.1103/PhysRevLett.55.2656}

\bibitem{Wheeler:1985qd}
J.~T.~Wheeler,
``Symmetric Solutions to the Maximally {Gauss-Bonnet} Extended 
Einstein Equations,''
Nucl.\ Phys.\ B {\bf 273} (1986) 732.
\doi{10.1016/0550-3213(86)90388-3}

\bibitem{Behrndt:1998eq}
K.~Behrndt, G.~Lopes Cardoso, B.~de Wit, D.~L\"ust, T.~Mohaupt and W.~A.~Sabra,
``Higher order black hole solutions in N=2 supergravity and 
Calabi-Yau string backgrounds,''
Phys.\ Lett.\ B {\bf 429} (1998) 289.
\doi{10.1016/S0370-2693(98)00413-4}
[\hepth{9801081}].

\bibitem{Cai:2001dz}
R.~G.~Cai,
``Gauss-Bonnet black holes in AdS spaces,''
Phys.\ Rev.\ D {\bf 65} (2002) 084014.
\doi{10.1103/PhysRevD.65.084014}
[\hepth{0109133}].

\bibitem{Oliva:2010eb}
J.~Oliva and S.~Ray,
``A new cubic theory of gravity in five dimensions: Black hole, B
irkhoff's theorem and C-function,''
Class.\ Quant.\ Grav.\  {\bf 27} (2010) 225002.
\doi{10.1088/0264-9381/27/22/225002}
[\arxiv{1003.4773} [gr-qc]].

\bibitem{Myers:2010ru}
R.~C.~Myers and B.~Robinson,
``Black Holes in Quasi-topological Gravity,''
JHEP {\bf 1008} (2010) 067.
\doi{10.1007/JHEP08(2010)067}
[\arxiv{1003.5357} [gr-qc]].

\bibitem{Lu:2015cqa}
H.~Lu, A.~Perkins, C.~N.~Pope and K.~S.~Stelle,
``Black Holes in Higher-Derivative Gravity,''
Phys.\ Rev.\ Lett.\  {\bf 114} (2015) no.17,  171601.
\doi{10.1103/PhysRevLett.114.171601}
[\arxiv{1502.01028} [hep-th]].

\bibitem{Kehagias:2015ata}
A.~Kehagias, C.~Kounnas, D.~L\"ust and A.~Riotto,
``Black hole solutions in $R^{2}$ gravity,''
JHEP {\bf 1505} (2015) 143.
\doi{10.1007/JHEP05(2015)143}
[\arxiv{1502.04192} [hep-th]].

\bibitem{Hendi:2015hba}
S.~H.~Hendi, S.~Panahiyan and B.~Eslam Panah,
``Magnetic branes in Gauss-Bonnet gravity with nonlinear electrodynamics: 
correction of magnetic branes in Einstein-Maxwell gravity,''
Eur.\ Phys.\ J.\ C {\bf 75} (2015) no.6,  296
\doi{10.1140/epjc/s10052-015-3521-7}
[\arxiv{1506.02481} [gr-qc]].


\bibitem{Hennigar:2016gkm}
R.~A.~Hennigar and R.~B.~Mann,
``Black holes in Einsteinian cubic gravity,''
Phys.\ Rev.\ D {\bf 95} (2017) no.6,  064055
\doi{10.1103/PhysRevD.95.064055}
[\arxiv{1610.06675} [hep-th]].

\bibitem{Bueno:2016lrh}
P.~Bueno and P.~A.~Cano,
``Four-dimensional black holes in Einsteinian cubic gravity,''
Phys.\ Rev.\ D {\bf 94} (2016) no.12,  124051.
\doi{10.1103/PhysRevD.94.124051}
[\arxiv{1610.08019} [hep-th]].

\bibitem{Hennigar:2017ego}
R.~A.~Hennigar, D.~Kubiznak and R.~B.~Mann,
``Generalized quasi-topological gravity,''
\arxiv{1703.01631} [hep-th].

\bibitem{Bueno:2017sui}
P.~Bueno and P.~A.~Cano,
``On black holes in higher-derivative gravities,''
\arxiv{1703.04625} [hep-th].

\bibitem{Olmo:2011uz}
G.~J.~Olmo,
``Palatini Approach to Modified Gravity: f(R) Theories and Beyond,''
Int.\ J.\ Mod.\ Phys.\ D {\bf 20} (2011) 413.
\doi{10.1142/S0218271811018925}.
[\arxiv{1101.3864} [gr-qc]].

\bibitem{Stelle:1976gc}
K.~S.~Stelle,
``Renormalization of Higher Derivative Quantum Gravity,''
Phys.\ Rev.\ D {\bf 16} (1977) 953.
\doi{10.1103/PhysRevD.16.953}

\bibitem{Stelle:1977ry}
K.~S.~Stelle,
``Classical Gravity with Higher Derivatives,''
Gen.\ Rel.\ Grav.\  {\bf 9} (1978) 353.
\doi{10.1007/BF00760427}

\bibitem{Julve:1978xn}
J.~Julve and M.~Tonin,
``Quantum Gravity with Higher Derivative Terms,''
Nuovo Cim.\ B {\bf 46} (1978) 137.
\doi{10.1007/BF02748637}

\bibitem{Boulware:1983td}
D.~G.~Boulware, G.~T.~Horowitz and A.~Strominger,
``Zero Energy Theorem for Scale Invariant Gravity,''
Phys.\ Rev.\ Lett.\  {\bf 50} (1983) 1726.
\doi{10.1103/PhysRevLett.50.1726}

\bibitem{David:1984uv}
F.~David and A.~Strominger,
``On the Calculability of Newton's Constant and the Renormalizability of Scale Invariant Quantum Gravity,''
Phys.\ Lett.\  {\bf 143B} (1984) 125.
\doi{10.1016/0370-2693(84)90817-7}

\bibitem{Horowitz:1984wv}
G.~T.~Horowitz,
``Quantum Cosmology With a Positive Definite Action,''
Phys.\ Rev.\ D {\bf 31} (1985) 1169.
\doi{10.1103/PhysRevD.31.1169}

\bibitem{Deser:2007vs}
S.~Deser and B.~Tekin,
``New energy definition for higher curvature gravities,''
Phys.\ Rev.\ D {\bf 75} (2007) 084032.
\doi{10.1103/PhysRevD.75.084032}
[\grqc{0701140}].

\bibitem{kn:Love1970}
D. Lovelock, 
``Divergence-free tensorial concomitants,''
\textit{Aequationes Math.} \textbf{4}  (1970) 127.

\bibitem{Lovelock:1971yv}
D.~Lovelock,
``The Einstein tensor and its generalizations,''
J.\ Math.\ Phys.\  {\bf 12} (1971) 498.
\doi{10.1063/1.1665613}

\bibitem{Bergshoeff:2009hq}
E.~A.~Bergshoeff, O.~Hohm and P.~K.~Townsend,
``Massive Gravity in Three Dimensions,''
Phys.\ Rev.\ Lett.\  {\bf 102} (2009) 201301.
\doi{10.1103/PhysRevLett.102.201301}
[\arxiv{0901.1766} [hep-th]].

\bibitem{Lu:2011zk}
H.~Lu and C.~N.~Pope,
``Critical Gravity in Four Dimensions,''
Phys.\ Rev.\ Lett.\  {\bf 106} (2011) 181302.
\doi{10.1103/PhysRevLett.106.181302}
[\arxiv{1101.1971} [hep-th]].

\bibitem{Karasu:2016ifk}
A.~Karasu, E.~Kenar and B.~Tekin,
``Minimal extension of Einstein's theory: The quartic gravity,''
Phys.\ Rev.\ D {\bf 93} (2016) no.8,  084040.
\doi{10.1103/PhysRevD.93.084040}
[\arxiv{1602.02567} [hep-th]].

\bibitem{Bueno:2016xff}
P.~Bueno and P.~A.~Cano,
``Einsteinian cubic gravity,''
Phys.\ Rev.\ D {\bf 94} (2016) no.10,  104005.
\doi{10.1103/PhysRevD.94.104005}
[\arxiv{1607.06463} [hep-th]].

\bibitem{Sisman:2011gz}
T.~C.~Sisman, I.~Gullu and B.~Tekin,
``All unitary cubic curvature gravities in D dimensions,''
Class.\ Quant.\ Grav.\  {\bf 28} (2011) 195004.
\doi{10.1088/0264-9381/28/19/195004}
[\arxiv{1103.2307} [hep-th]].

\bibitem{Tekin:2016vli}
B.~Tekin,
``Particle Content of Quadratic and $f(R_{\mu\nu\sigma \rho})$ 
Theories in $(A)dS$,''
Phys.\ Rev.\ D {\bf 93} (2016) no.10,  101502.
\doi{10.1103/PhysRevD.93.101502}
[\arxiv{1604.00891} [hep-th]].

\bibitem{Bueno:2016ypa}
P.~Bueno, P.~A.~Cano, V.~S.~Min and M.~R.~Visser,
``Aspects of general higher-order gravities,''
Phys.\ Rev.\ D {\bf 95} (2017) no.4,  044010.
\doi{10.1103/PhysRevD.95.044010}
[\arxiv{1610.08519} [hep-th]].

\bibitem{Sotiriou:2008rp}
T.~P.~Sotiriou and V.~Faraoni,
``f(R) Theories Of Gravity,''
Rev.\ Mod.\ Phys.\  {\bf 82} (2010) 451.
\doi{10.1103/RevModPhys.82.451}
[\arxiv{0805.1726} [gr-qc]].

\bibitem{DeFelice:2010aj}
A.~De Felice and S.~Tsujikawa,
``f(R) theories,''
Living Rev.\ Rel.\  {\bf 13} (2010) 3.
\doi{10.12942/lrr-2010-3}
[\arxiv{1002.4928} [gr-qc]].

\bibitem{Nojiri:2010wj}
S.~Nojiri and S.~D.~Odintsov,
``Unified cosmic history in modified gravity: from F(R) 
theory to Lorentz non-invariant models,''
Phys.\ Rept.\ {\bf 505} (2011) 59.
\doi{10.1016/j.physrep.2011.04.001}
[\arxiv{1011.0544} [gr-qc]].

\bibitem{Sarkar:2013swa}
S.~Sarkar and A.~C.~Wall,
``Generalized second law at linear order for actions that are fun
ctions of Lovelock densities,''
Phys.\ Rev.\ D {\bf 88} (2013) 044017.
\doi{10.1103/PhysRevD.88.044017}
[\arxiv{1306.1623} [gr-qc]].

\bibitem{Bueno:2016dol}
P.~Bueno, P.~A.~Cano, A.~O.~Lasso and P.~F.~Ram\'{\i}rez,
``f(Lovelock) theories of gravity,''
JHEP {\bf 1604} (2016) 028.
\doi{10.1007/JHEP04(2016)028}
[\arxiv{1602.07310} [hep-th]].

\bibitem{kn:BG}
B.~Bertotti and L.~P.~Grishchuk,
``The strong equivalence principle,''
Class.\ Quant.\ Grav.\  {\bf 7} (1990) 1733-1745.

\bibitem{Will:2014kxa}
C.~M.~Will,
``The Confrontation between General Relativity and Experiment,''
Living Rev.\ Rel.\  {\bf 17} (2014) 4.
\doi{10.12942/lrr-2014-4}
[\arxiv{1403.7377} [gr-qc]].

\bibitem{DiCasola:2013iia}
E.~Di Casola, S.~Liberati and S.~Sonego,
``Nonequivalence of equivalence principles,''
Am.\ J.\ Phys.\  {\bf 83} (2015) 39.
\doi{10.1119/1.4895342}
[\arxiv{1310.7426} [gr-qc]].

\bibitem{Ortin:2015hya}
T.~Ort\'{\i}n,
``Gravity and Strings'', 2$^{\rm nd}$ edition
Cambridge University Press (2015)

\bibitem{Weinberg:1964ew}
S.~Weinberg,
``Photons and Gravitons in s Matrix Theory: Derivation of Charge 
Conservation and Equality of Gravitational and Inertial Mass,''
Phys.\ Rev.\  {\bf 135} (1964) B1049.
\doi{10.1103/PhysRev.135.B1049}

\bibitem{kn:We3} 
S.~Weinberg,
``Derivation of Gauge Invariance and the 
  Equivalence Principle from Lorentz Invariance
in the S Matrix,''
Phys.\ Lett.\ {\bf 9} (1964) 357--359.

\bibitem{Boulware:1974sr}
D.~G.~Boulware and S.~Deser,
``Classical General Relativity Derived from Quantum Gravity,''
Annals Phys.\  {\bf 89} (1975) 193.
\doi{10.1016/0003-4916(75)90302-4}

\bibitem{Gupta:1954zz}
S.~N.~Gupta,
``Gravitation and Electromagnetism,''
Phys.\ Rev.\  {\bf 96} (1954) 1683.
\doi{10.1103/PhysRev.96.1683}

\bibitem{kn:Gup2} 
S.\,N.~Gupta,
``Einstein's and Other Theories of Gravitation,''
Rev.\ Mod.\ Phys.\ {\bf 29} (1957) 334--336.

\bibitem{Kraichnan:1955zz}
R.~H.~Kraichnan,
``Special-Relativistic Derivation of Generally Covariant Gravitation Theory,''
Phys.\ Rev.\  {\bf 98} (1955) 1118.
\doi{10.1103/PhysRev.98.1118}

\bibitem{Deser:1969wk}
S.~Deser,
``Selfinteraction and gauge invariance,''
Gen.\ Rel.\ Grav.\  {\bf 1} (1970) 9
\doi{10.1007/BF00759198}
[\grqc{0411023}].

\bibitem{Butcher:2009ta}
L.~M.~Butcher, M.~Hobson and A.~Lasenby,
``Bootstrapping gravity: A Consistent approach to 
energy-momentum self-coupling,''
Phys.\ Rev.\ D {\bf 80} (2009) 084014.
\doi{10.1103/PhysRevD.80.084014}
[\arxiv{0906.0926} [gr-qc]].

\bibitem{Deser:2009fq}
S.~Deser,
``Gravity from self-interaction redux,''
Gen.\ Rel.\ Grav.\  {\bf 42} (2010) 641.
\doi{10.1007/s10714-009-0912-9}
[\arxiv{0910.2975} [gr-qc]].

\bibitem{Barcelo:2014mua}
C.~Barcel\'o, R.~Carballo-Rubio and L.~J.~Garay,
``Unimodular gravity and general relativity from graviton self-interactions,''
Phys.\ Rev.\ D {\bf 89} (2014) no.12,  124019.
\doi{10.1103/PhysRevD.89.124019}
[\arxiv{1401.2941} [gr-qc]].

\bibitem{Nordtvedt:1968qr}
K.~Nordtvedt,
``Equivalence Principle for Massive Bodies. 1. Phenomenology,''
Phys.\ Rev.\  {\bf 169} (1968) 1014.
\doi{10.1103/PhysRev.169.1014}

\bibitem{Nordtvedt:1968qs}
K.~Nordtvedt,
``Equivalence Principle for Massive Bodies. 2. Theory,''
Phys.\ Rev.\  {\bf 169} (1968) 1017.
\doi{10.1103/PhysRev.169.1017}

\bibitem{kn:Kra2} 
R.\,H.~Kraichnan,
``Possibility of Unequal Gravitational and Inertial Masses,''
Phys.\ Rev.\ {\bf 107} (1957) 1485--1490.

\bibitem{kn:Ros} 
L.~Rosenfeld,
``Sur le tenseur d'impulsion-\'energie,''
M\'em.\ Acad.\ Roy.\ Belgique {\bf 6} (1930) 30.

\bibitem{kn:Bel} 
F.~J.~Belinfante,
``On the Spin Angular Momentum of Mesons,''
Physica \textbf{VI} (1939) 887.

\bibitem{Feynman:1996kb}
R.~P.~Feynman, F.~B.~Morinigo, W.~G.~Wagner and B.~Hatfield,
``Feynman lectures on gravitation,''
Reading, USA: Addison-Wesley (1995).

\bibitem{kn:Thir2} 
W.~E.~Thirring,
``An Alternative Approach to the Theory of Gravitation,''
Ann.\ Phys.\ \textbf{16} (1964) 96--117.

\bibitem{Kreuzer:1989ec}
M.~Kreuzer,
``Gauge Theory of Volume Preserving Diffeomorphisms,''
Class.\ Quant.\ Grav.\  {\bf 7} (1990) 1303.
\doi{10.1088/0264-9381/7/8/010}

\bibitem{Alvarez:2005iy}
E.~\'Alvarez,
``Can one tell Einstein's unimodular theory from Einstein's general relativity?,''
JHEP {\bf 0503} (2005) 002.
\doi{10.1088/1126-6708/2005/03/002}
[\hepth{0501146}].

\bibitem{Alvarez:2006uu}
E.~\'Alvarez, D.~Blas, J.~Garriga and E.~Verdaguer,
``Transverse Fierz-Pauli symmetry,''
Nucl.\ Phys.\ B {\bf 756} (2006) 148.
\doi{10.1016/j.nuclphysb.2006.08.003}
[\hepth{0606019}].

\bibitem{kn:Eins}
A.~Einstein, (1919) Siz. Preuss. Acad. Scis., english translation in ``The
principle of relativity'', by A. Einstein et al. (Dover).

\bibitem{Deser:1987uk}
S.~Deser,
``Gravity From Selfinteraction in a Curved Background,''
Class.\ Quant.\ Grav.\  {\bf 4} (1987) L99.
\doi{10.1088/0264-9381/4/4/006}

\end{thebibliography}
\end{document}